# Dislocation Transmission Across Tilt Low-Angle Grain Boundaries in BCC Fe: The Role of Elastic Interactions

Shuai Zhang[1], Zhishun Chen[1], Zhuoming Xie[2], Jun Song[3], Huiqiu Deng[4], Wangyu Hu[1,*], Jie Hou[1,*]

1. College of Materials Science and Engineering, State Key Laboratory of Cemented Carbide, Hunan University, Changsha 410082, China
2. Key Laboratory of Materials Physics, Institute of Solid State Physics, Chinese Academy of Sciences, P. O. Box 1129, Hefei 230031, PR China
3. Department of Mining and Materials Engineering, McGill University, Montreal, Quebec, H3A 0C5, Canada
4. School of Physics and Electronics, Hunan University, Changsha 410082, China

[*] Corresponding authors: wyuhu@hnu.edu.cn (Wangyu Hu), jiehou@hnu.edu.cn (Jie Hou)

## Abstract

Low-angle grain boundaries (LAGBs) are often regarded as penetrable interfaces to dislocation motion, yet recent studies suggest they can also act as strong barriers. The origin of this duality remains debated, particularly regarding the role of elastic interactions. Here, large-scale molecular dynamics simulations are employed to investigate dislocation transmission across various tilt LAGBs in BCC Fe. The results show that transmission resistance varies widely with boundary-dislocation geometry. Contrary to the prevailing view that dislocation reactions dominate, elastic interactions between lattice and boundary dislocations emerge as the primary controlling factor. Screw and screw-like dislocations generate shear stresses that bend GB dislocations and produce strong barriers, whereas edge dislocations lack such stresses and transmit more readily. Consequently, barrier strength increases as the dislocation character angle decreases, with screw dislocations experiencing the strongest resistance. From these insights, we develop an analytical model that quantitatively links net transmission stress to dislocation character, boundary inclination, and boundary misorientation, reproducing the simulation results with excellent agreement. These results establish the dominant role of elastic interactions in dislocation-LAGB interactions and provide a predictive basis for designing materials strengthened by controlled boundary architectures.

**Key words**: low-angle grain boundary; dislocation; molecular dynamics; iron and steel;

# 1. Introduction

Dislocation-grain boundary (GB) interactions play a central role in governing the strength, plasticity, and fracture behavior of crystalline materials[1-6]. Acting as barriers to dislocation glide, GBs control plastic strain accommodation and GB-mediated strengthening, with their interaction behavior strongly dependent on GB structure. While dislocation transmission and absorption at high-angle grain boundaries (HAGBs)[3-8] have been extensively investigated, far less is known about their behavior at low-angle grain boundaries (LAGBs), leaving the role of LAGBs in dislocation transmission largely unresolved and subject to ongoing debated.

LAGBs, comprising ordered arrays of dislocations[9], can be introduced into materials through dynamic plastic deformation combined with appropriate annealing[10, 11], mild dynamic recovery[12], or thermal treatments[13, 14]. Prevalent in a wide range of metallic system (e.g., lath martensitic steels[13, 14]), LAGBs known to influence strength, ductility[10-12], and resistance to fatigue [2, 3] and stress corrosion cracking[10, 11]. Their impact arises largely from interactions with dislocations, which depend sensitively on whether these boundaries are penetrable[2, 3, 15-20]. HAGBs are generally regarded as strong barriers that promote dislocation pile-ups and stress concentration, making them preferential crack initiation sites[1-8, 15-18]. In contrast, LAGBs are often considered weak obstacles due to their low misorientations and discrete dislocation structures[2, 3, 12, 17], and are therefore assumed to offer minimal resistance to dislocation motion, consistent with the absence of fatigue cracking observed along such boundaries[2, 3].

Despite the long-standing view of LAGBs as weak and largely transparent interfaces, recent findings demonstrates that under certain conditions, LAGBs can impede dislocation motion, concentrate local stresses, and even act as preferential sites for crack initiation and propagation[15, 16, 18-24]. Nagao et al.[15, 16, 18] observed "quasi-cleavage" fractures along LAGBs in hydrogen-charged martensitic steels, suggesting that LAGBs can block dislocations and induce stress concentration. In situ nanoindentation in $SrTiO_3$ bicrystals by Kondo et al.[23] indicated that although dislocations generally transmit more readily across LAGBs than HAGBs, local boundary stabilization can still impede dislocation motion. Ueki et al.[20] observed a pronounced dependence on dislocation character, with in-habit-plane edge dislocations transmitting across LAGB readily, whereas out-of-habit-plane dislocations that dominated by screw components require much higher stresses for transmission. Chen et al.[19] further demonstrated that LAGBs behave as "tunable" obstacles, allowing orientation dependent selective transmission of dislocations through LAGBs. These experimental point to a possible dual nature of LAGBs, as either penetrable interfaces or effective barriers of dislocations. However, due to experimental challenges in atomistic scale characterization, current insights remain fragmented, and the underlying mechanisms governing dislocation-LAGB interactions remains elusive.

To overcome these challenges, numerical simulations have been employed to reveal fundamentals of LAGB-dislocation interactions. Using discrete dislocation dynamics (DDD), Liu et al.[21, 22] demonstrated that boundary dislocations in LAGBs significantly impede the transmission of lattice dislocations in $\alpha$-Fe, leading to junction formation and significant blocking. Transmission was governed by the stability of these junctions, including binary <001> junctions, collinear annihilation, and ternary 1/2<111> junctions, with the latter two often rendering LAGBs nearly impenetrable. Similar conclusions have been reported for FCC metals[25, 26], emphasizing the dominant role of dislocation reaction and junction formation. Nevertheless, DDD remains inherently limited by its coarse-grained frameworks and empirical parameter dependence, precluding access to local atomic structural evolution at the boundary.

Atomistic simulations offer a fundamental perspective by directly capturing the atomic-scale mechanisms that govern dislocation-LAGB interactions. Gao et al.[27] conducted molecular dynamics (MD) simulations of screw and mixed dislocations interacting with <110> tilt Lomer-type LAGBs in FCC Ni, identifying complex dislocation reactions as the dominant blocking mechanism. However, their analysis was confined to a special case where boundary and lattice dislocations were parallel, often resulting in nearly unhindered transmission. Wakeda et al.[24] utilized nanocrystalline models with artificially inserted dislocation loops to investigate interactions in α-Fe, demonstrating stronger blocking of screw dislocations compared to edge segments. They attribute this to large line tension of the screw component, a conclusion that contradicts established energetics in BCC metals, where edge dislocations typically possess greater line energy[28]. Although atomistic approaches are uniquely capable of capturing these interactions in detail, current studies remain narrow scoped and focused primarily on dislocation reactions[21, 22, 24-27], largely neglecting the elastic interactions between incoming dislocations and the LAGB network—an aspect that may be critical to understanding the fundamental origin of the LAGB barrier effect.

To address this gap, we systematically investigated the interactions between various lattice dislocations and well-defined tilt LAGBs in BCC Fe using MD simulations. Our results demonstrate that the elastic interaction between the incoming dislocation and boundary dislocations dominates the transmission resistance by inducing pronounced bending of boundary dislocations, whereas dislocation reactions that emphasized in previous studies play only a secondary role. This elastic interaction is strongly dependent on dislocation character, with near-screw dislocations causing greater bending and thus requiring higher critical resolved shear stress (CRSS) for transmission, agreeing well with recent experimental observations. Furthermore, we establish an analytical model that quantitatively links net transmission stress to dislocation character, boundary inclination, and boundary misorientation, reproducing the simulation results with excellent agreement. These findings underscore the crucial role of elastic interactions in governing dislocation-LAGB behavior and provide new insight into the fundamental barrier mechanisms of LAGBs.

## 2. Computational methods

All simulations in this study were performed using the Large-scale Atomic/Molecular Massively Parallel Simulator (LAMMPS) package[29]. Structural analysis and visualization were conducted using OVITO[30], which integrates the Dislocation Extraction Algorithm (DXA)[31] method to identify dislocations. In our MD simulations, the interatomic interactions between Fe atoms were described using the potential developed by Ackland et al.[32], which is an optimized version of Potential 2 by Mendelev et al.[33], modified with additional point defect data. The potential employed in our simulations was fitted using data from perfect crystals, point defects, and density functional theory (DFT) forces, and it accurately reproduces screw dislocation core structures consistent with DFT calculations. Model construction was facilitated by the Atomsk software package[34] and the dislocation embedding code developed by Zhang et al.[35]. Finite temperature MD simulations were performed using the Nosé-Hoover thermostat with a time step of 1 fs.

## 2.1. Construction of LAGB models

In this work, we constructed bi-crystal models containing tilt LAGBs lying along $(1\bar{1}0)$, $(111)$, $(010)$ crystalline planes, with tilt axes aligned to low-index crystallographic directions. As summarized in **Table 1**, eight distinct LAGB types were investigated, each with five misorientation angles ($\theta$) ranging from ~1° to ~5° adopted for each LAGB type. Minor fluctuations in $\theta$ arise from the discrete nature of atomic positions. This range is representative of typical misorientations observed in lath boundaries of martensitic steels[14].

To identify the most stable LAGB configurations, we employed two complementary construction methods (**Figs. 1a-1b**). In method A (**Fig. 1a**), the upper and lower grains were rotated by $\pm\theta/2$ about the tilt axis while maintaining periodicity along the Z direction. The top grain was laterally shifted by a variable amount $\Delta l$ before merging with the bottom grain to form a bi-crystal. The resulting structure was gradually annealed from 1200 K to 0 K, followed by energy minimizations. The GB energy was computed as:

$$E_{GB} = \frac{E_{mid} - E_{bulk}N_{mid}}{S}, \qquad (1)$$

where $E_{mid}$ is the total energy of atoms in a GB-centered region (excluding regions within 20 Å from free surfaces, the thickness of this region is at least 1600 Å along Y direction to ensure negligible surface-LAGB interaction); $N_{mid}$ is the number of atoms in this region; $E_{bulk}$ is the per-atom energy of perfect bulk Fe, and S is the GB area. For each misorientation, the $\Delta l$ yielding the lowest $E_{GB}$ was selected.

**Table 1.** Configurational parameters of the LAGB-dislocation interaction model. GB type indicates the grain boundary plane and tilt axis. $\overrightarrow{b_{fr}}$ denotes Burgers vector of the incoming free dislocation. X, Y, and Z correspond to the simulation box directions as illustrated in **Fig. 1c**, with X and Z aligned with the slip plane normal and the line direction of the free dislocation, respectively. α is the character angle of the free dislocation, and β is the angle between the LAGB plane and the slip plane of the free dislocation. Each configuration is labeled as "Notation±" (e.g., A1+, B2-), where the sign indicates the direction of $\overrightarrow{b_{fr}}$.

| LAGB | $\overrightarrow{b_{fr}}$ | X | Y | Z | α(°) | β(°) | Notation |
|---|---|---|---|---|---|---|---|
| $(1\bar{1}0)\,[001]$ | | $[\bar{1}12]$ | $[1\bar{1}1]$ | $[110]$ | 90.00 | 54.74 | A1 |
| $(1\bar{1}0)\,[110]$ | $\pm 1/2[1\bar{1}1]$ | $[110]$ | $[1\bar{1}0]$ | $[00\bar{1}]$ | 54.74 | 90.00 | A2 |
| $(1\bar{1}0)\,[11\bar{1}]$ | | $[31\bar{2}]$ | $[2\bar{4}1]$ | $[\bar{1}\bar{1}2]$ | 61.87 | 67.79 | A3 |
| $(1\bar{1}0)\,[11\bar{2}]$ | | $[\bar{1}01]$ | $[\bar{1}2\bar{1}]$ | $[\bar{1}\bar{1}\bar{1}]$ | 70.53 | 60.00 | A4 |
| $(111)\,[1\bar{1}0]$ | $\pm 1/2[11\bar{1}]$ | $[1\bar{1}0]$ | $[111]$ | $[\bar{1}\bar{1}2]$ | 19.47 | 90.00 | B1 |
| | $\pm 1/2[\bar{1}11]$ | $[312]$ | $[\bar{2}41]$ | | 61.87 | 22.21 | B2 |
| $(111)\,[11\bar{2}]$ | $\pm 1/2[11\bar{1}]$ | $[112]$ | $[\bar{1}\bar{1}1]$ | $[1\bar{1}0]$ | 90.00 | 19.47 | B3 |
| | $\pm 1/2[\bar{1}11]$ | $[110]$ | $[001]$ | | 35.26 | 35.26 | B4 |
| $(010)\,[100]$ | $\pm 1/2[111]$ | $[1\bar{1}0]$ | $[110]$ | $[001]$ | 54.74 | 45.00 | C1 |
| $(010)\,[101]$ | $\pm 1/2[11\bar{1}]$ | $[101]$ | $[010]$ | $[\bar{1}01]$ | 35.26 | 90.00 | C2 |
| | $\pm 1/2[111]$ | $[1\bar{2}1]$ | $[111]$ | | 90.00 | 35.26 | C3 |

While method A consistently produced uniformly spaced <111>/2 and <010> edge dislocations for LAGBs along (111) and (010) planes, it often failed for boundaries on $(1\bar{1}0)$, yielding unnaturally paired $<1\bar{1}1>/2$ and $<1\bar{1}\bar{1}>/2$ dislocations with anomalously high GB energies. To overcome this, we adopted method B (**Fig. 1b**), where dislocations with Burgers vectors $b_1=<1\bar{1}1>/2$ and $b_2=<1\bar{1}\bar{1}>/2$ were manually inserted at controlled spacing $d$, and the final box length along Z was set to $l_z = \frac{|b_1+b_2|}{2\sin(\theta/2)}$[9]. Note dislocation insertion often disrupted atomic continuity at periodic boundaries, so atoms in buffer zones (purple in **Fig. 1b**) were removed and boundaries adjusted to restore periodicity. Subsequent annealing and energy minimization were performed identically to method A. Across all cases, the lowest-energy structures comprised regularly spaced dislocations, in agreement with predictions from dislocation theory[9].

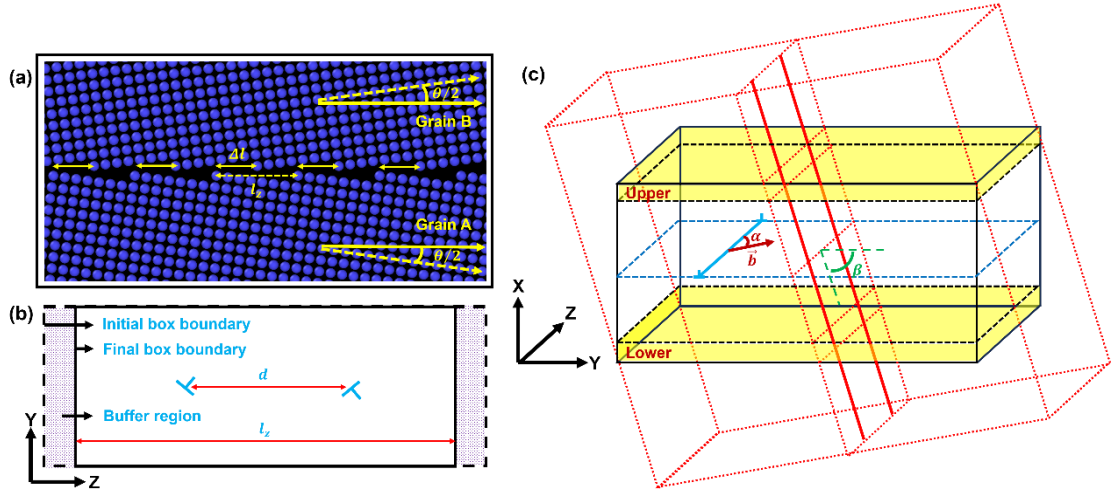

**Fig. 1.** Construction of atomistic models. (a) Schematic of LAGB generation via method A, where two grains are merged after relative rotation and lateral shifting. (b) Construction via method B by direct dislocation insertion; the buffer region (purple) has a thickness equal to half lattice period of BCC Fe along the corresponding direction and is removed to restore periodicity. (c) Simulation setup for dislocation-LAGB interaction. The red dashed box indicates the initial LAGB model with boundary dislocations (red lines); the black solid box shows the final model after rotation and truncation. Fixed layers (yellow) are used to apply shear. The free dislocation (solid blue line) is characterized by Burgers vector $\vec{b}$ and character angle $\alpha$. $\beta$ is the angle between the LAGB and the slip plane (blue dashed). The simulation box is periodic along Z direction.

## 2.2. Construction of dislocation-LAGB interaction models

After identifying the most stable LAGB structures, we constructed models to examine their interactions with a nearby free dislocation at misorientation angles of ~2° and ~4°. Each model contained two GB dislocations (red lines, **Fig. 1c**) and was rotated to align the YZ plane with the slip plane of the incoming dislocation (blue dashed). The simulation box was truncated into an orthogonal shape (black box). A free dislocation (solid blue line) was then inserted approximately 1 nm from the LAGB using the method of Zhang et al.[35]. Periodic boundary conditions were applied along Z to represent an infinitely long dislocation, while X and Y were non-periodic. The X dimension was set to 22 nm, including rigid layers (0.8 nm) exceeding the interatomic potential cutoff. The Y dimension, ranging from 56 to 66 nm depending on the inclination angle β between the LAGB and slip plane, ensured that the GB plane intersected only the top and bottom surfaces. To restore periodicity along Z that disrupted by dislocation insertion, a layer of atoms near the Z boundary was replaced with bulk configurations, and overlapping atoms within a cutoff distance were removed, yielding a final Z length of two times the GB dislocation spacing to accommodate two GB dislocations.

The detailed model parameters are summarized in **Table 1**. A few principles for constructing the dislocation-LAGB interaction model are worth noting. First, only free dislocations with Burgers vectors that are not parallel to the GB plane were included, as parallel dislocations do not intersect the LAGB during slipping. Second, for $(1\bar{1}0)$ LAGBs, this condition restricts the Burgers vector to $\pm 1/2[1\bar{1}1]$ or $\pm 1/2[1\bar{1}\bar{1}]$, which align with one of the alternating $1/2[1\bar{1}1]$ and $1/2[1\bar{1}\bar{1}]$ dislocations constituting the LAGB. Third, we focus on {110}, {112}, and {123} slip planes for free dislocations due to their relatively high mobility[36]; symmetrically equivalent configurations were not explicitly explored, and only ~2° and ~4° LAGBs were examined owing to computational cost constraints. In total, 44 distinct dislocation-LAGB interaction models were constructed. Finally, pure screw-type free dislocations ($\alpha = 0$) were not considered in this study, as their transmission through LAGBs under lattice misorientation inevitably induces jog formation. These jogs produce vacancy trails during subsequent glide due to jog dragging within the constraints of periodic boundaries, introducing extrinsic effects beyond the scope of this work.

## 2.3. Relaxation and loading of dislocation-LAGB interaction models

To prevent absorption of GB dislocations by the surfaces along the X direction, atoms in the Upper and Lower regions (**Fig. 1c**) were rigidly fixed throughout both relaxation and loading. For each dislocation-LAGB model, five independent simulations with different random initial velocities were conducted to reduce the influence of thermal fluctuations.

Before loading, a sequential relaxation procedure was performed to equilibrate the system. The models were first energy-minimized at 0 K using the conjugate gradient algorithm, then thermally scaled to 300 K. This was followed by isothermal-isobaric (NPT with P=0) equilibration at 300 K for 100 ps. To satisfy the NPT conditions in LAMMPS, temporary vacuum layers were introduced along the non-periodic X and Y directions and removed prior to loading. This relaxation ensured stable configurations with adequate interaction between the free and GB dislocations.

Shear loading was applied under the NPT ensemble by displacing the Upper layer along the Burgers vector direction of the free dislocation at a shear strain rate of 0.05/ns, while keeping the Lower layer fixed. The principal stress along the periodic Z direction was maintained close to zero throughout loading.

## 3. Results

### 3.1. Structures and energies of LAGBs

We began by examining the atomistic structures of LAGBs in BCC Fe. As summarized in **Table 1**, eight groups of LAGBs were constructed with GB planes along $(1\bar{1}0)$, (111), (010), and with various tilt axes. Each group contains five misorientation angles ranging from ~1° to ~5°. Using the structure search strategy detailed in **Section 2.1**, we identified the most stable configurations for each LAGB group. Representative ~4° LAGBs are shown in **Fig. 2** to illustrate their characteristic structures. For (1$\bar{1}$0) LAGBs (**Figs. 2a-2d**), all four models consist of periodic arrays of alternating 1/2[1$\bar{1}$1] and 1/2[1$\bar{1}\bar{1}$] dislocations, equivalent to edge-type [1$\bar{1}$0] super-dislocations with net Burgers vectors normal to the GB plane. While these LAGBs share a similar pattern of alternating dislocation arrays, they differ in their slip planes and in the character angles of the constituent dislocations. In (111) LAGBs (**Figs. 2e-2f**), 1/2[111] edge dislocations are periodically arranged, with ($\bar{1}\bar{1}$2) slip planes for [1$\bar{1}$0]-tilt and (1$\bar{1}$0) for [11$\bar{2}$]-tilt boundaries. Similarly, (010) LAGBs (**Figs. 2g-2h**) comprise arrays of [010] edge dislocations, located on (001) slip planes for [100]-tilt and ($\bar{1}$01) slip planes for [101]-tilt boundaries.

Overall, tilt LAGBs with these three low-index GB planes can generally be described as periodic arrays of edge dislocations with Burgers vectors normal to the GB plane: [1$\bar{1}$0] super-dislocations for (1$\bar{1}$0) LAGBs, 1/2[111] dislocations for (111) LAGBs, and [010] dislocations for (010) LAGBs. Notably, while both <110> and <001> dislocations in BCC crystals can be decomposed into two 1/2<111> dislocations, the (1$\bar{1}$0) LAGBs accommodate <110> dislocations as alternating 1/2<111> components due to their instability, whereas (010) LAGBs preserve stable <001> dislocations, in agreement with Frank's rule[9]. These observations confirm that these LAGBs show structures consistent with classical dislocation theories, validating their rationality.

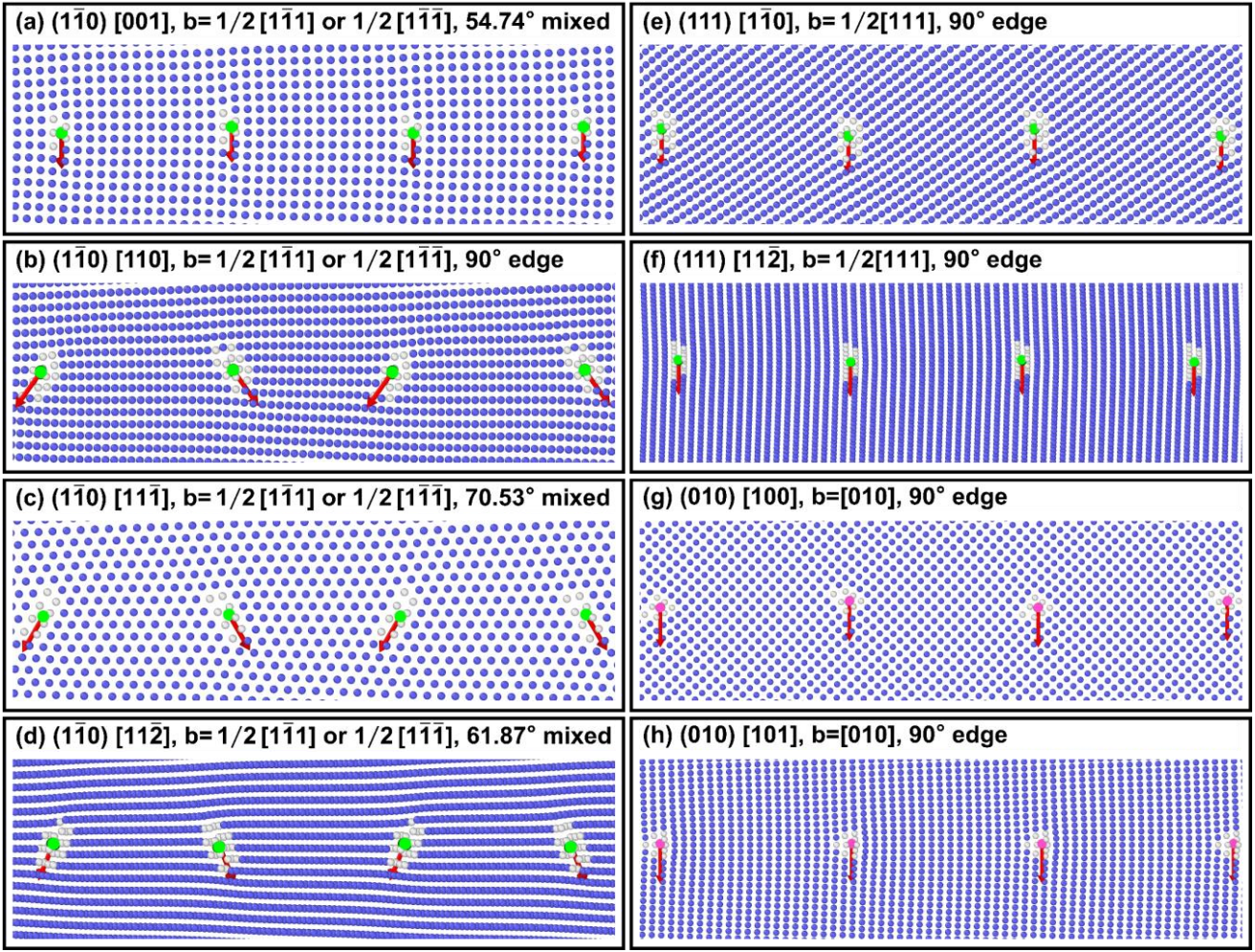

**Fig. 2.** Atomic structures of eight ~4° LAGBs. Blue atoms denote atoms in a perfect BCC lattice, whereas white atoms correspond to defective atoms. Dislocations identified by DXA[31] are shown as green (1/2<111>) and pink (<001>) lines, with red arrows indicating their Burgers vector directions. (a-d) (1$\bar{1}$0) LAGBs; (e-f) (111) LAGBs; (g-h) (010) LAGBs. Each panel is annotated with the GB plane and tilt axis, Burgers vector, character angle, and dislocation character (edge, screw, or mixed). The same color scheme for dislocations is used in all subsequent figures unless otherwise noted.

To assess the energetic stability of the constructed LAGBs, we further computed the GB energies of all eight LAGB types as a function of tilt angle, as shown in **Fig. 3**. The results are fitted using the Read-Shockley equation:

$$E_{GB} = E_0 \theta (A - \ln\theta), \tag{2}$$

where $E_0$ depends on the elastic properties of the material, and A is associated with core energy of an individual dislocation. The fitted parameters are summarized in **Table 2**. Across the investigated misorientation range, the GB energies increase with tilt angle, while the rate of increase decreases progressively, consistent with the Read-Shockley prediction. The excellent agreement between simulation data and these fittings strongly supports the structural reliability and physical validity of the constructed LAGBs.

The energy hierarchy among different LAGBs is primarily governed by the elastic strain energy of their constituent dislocations, as reflected by the fitted $E_0$ values. $(1\bar{1}0)$ LAGBs exhibit the highest energies, followed by (010) and then (111) LAGBs. This trend reflects the general energetic order of dislocations in BCC metals, where <110> dislocations are more energetically costly than <001> and 1/2<111> dislocations. Within each GB family, <110>-tilt LAGBs generally exhibit highest energies, whereas <001>-tilt LAGBs show lowest, consistent with the sequence of $E_0$ values in **Table 2**. While elastic strain energy dominates the overall energy trend, dislocation core energy parameter A induces local deviations from this general sequence. For example, the $(1\bar{1}0)[11\bar{1}]$ LAGB shows a slightly lower GB energy than a $(1\bar{1}0)[11\bar{2}]$ LAGB despite a larger $E_0$, due to its smaller A. A similar reversal occurs for (010)[101] LAGBs. These nuances highlight the competing roles of long-range strain fields and short-range core structures in governing GB energetics.

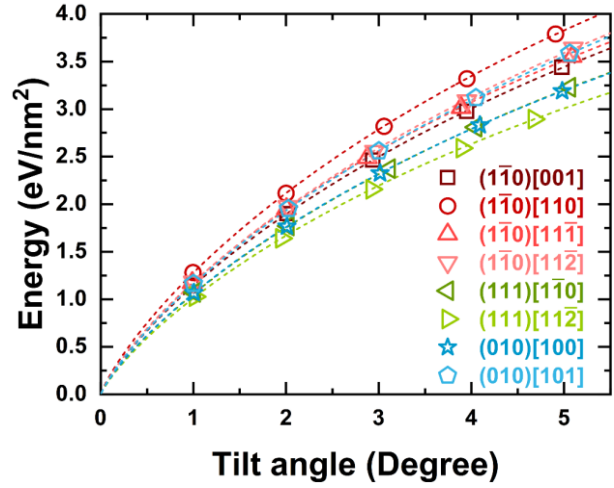

**Fig. 3.** GB energy as a function of tilt angle for eight types of LAGBs. Open symbols represent MD simulation results, and dashed lines correspond to fittings using the Read-Shockley equation (**Eq. (2)**).

**Table 2.** Fitting parameters $E_0$ and $A$ derived from Read-Shockley fits to the GB energy data shown in **Fig. 3**.

| LAGBs | $E_0(eV/nm^2)$ | A |
|---|---|---|
| $(1\bar{1}0)$ [001] | 16.165 | 0.0017 |
| $(1\bar{1}0)$ [110] | 18.307 | -0.0432 |
| $(1\bar{1}0)$ [11$\bar{1}$] | 16.966 | -0.0675 |
| $(1\bar{1}0)$ [11$\bar{2}$] | 16.321 | 0.0813 |
| (111) [1$\bar{1}$0] | 14.796 | 0.0401 |
| (111) [11$\bar{2}$] | 14.607 | -0.0799 |
| (010) [100] | 15.002 | 0.0024 |
| (010) [101] | 15.949 | 0.1166 |

## 3.2 Interaction and transmission of dislocations across LAGBs

After establishing the stable structures and energetics of pristine LAGBs, we introduced straight free dislocations near the LAGB models and performed energy minimization, zero-stress NPT relaxation, and subsequent shear loading, thereby capturing the intrinsic structural response and transmission behavior of dislocation across LAGBs. **Figs. 4a1-4c1** presents representative stress-strain curves together with the associated dislocation evolution during transmission across ~2° LAGBs on (1$\bar{1}$0), (111), and (010) planes, which correspond to interaction types A1+, B1+, and C2+ in **Table 1**, respectively. We first examined the dislocation structure right after the zero-stress NPT relaxation (**Figs. 4a2-4c2**), which provides a clean setting to dissect the elementary processes of dislocation binding, junction formation, and GB dislocation restructuring. We find that free dislocations spontaneously bind to LAGBs, giving rise to a variety of dislocation reactions, including collinear annihilation, the formation of binary <001> junctions, and the formation of ternary 1/2<111> junctions. For example, 1/2<111> free dislocations interacting with (1$\bar{1}$0) LAGBs form a <001> junction with one GB dislocation and undergo collinear annihilation with the other GB dislocation, whereas reactions with (111) LAGBs produce two binary <001> junctions. In contrast, interactions with (010) LAGBs primarily lead to two ternary 1/2<111> junctions involving <001> GB dislocations (a more detailed three-view projections of these configurations are provided in **Supplementary Fig. S1**). These dislocation reactions are broadly consistent with classical dislocation theory.

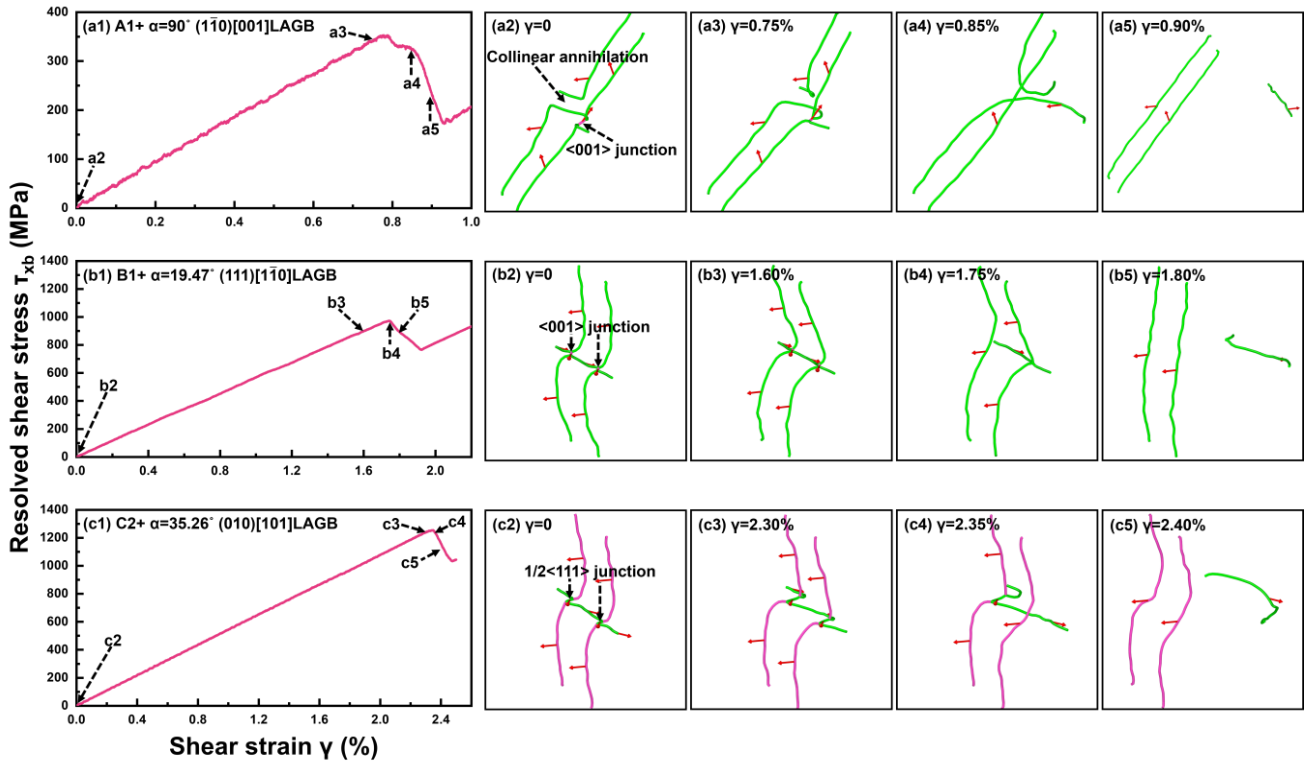

**Fig. 4.** Representative stress-strain responses and dislocation configurations illustrating the transmission of free dislocations across ~2° LAGBs. Three interaction types, (a1-a5) A1+, (b1-b5)

B1+, and (c1-c5) C2+, as defined in **Table 1**, are shown as typical cases. Four characteristic points are marked on each stress-strain curve to indicate characteristic stages, with corresponding dislocation configurations shown to the right of each curve. The dislocation color scheme follows **Fig. 2**, and the curves plot the resolved shear stress $\tau_{xb}$ versus strain, where x is the slip-plane normal and b is the Burgers vector of the free dislocation.

Following the zero-stress relaxation, we applied shear strain along the Burgers vector direction of the free dislocations to drive their transmission across LAGBs. As shown in **Figs. 4a1-4c1**, despite the distinct interaction mechanisms, all stress-strain responses exhibit three typical stages: (i) an initial elastic stage with linear stress increase, (ii) a transmission stage characterized by stress drop as the free dislocation transmits across the LAGB and glides toward the free surface, and (iii) a final elastic stage after the dislocation is absorbed by the free surface. The initial elastic stage primarily accumulates the resolved shear stress required to overcome the LAGB barriers for free dislocations. Stress rises linearly until reaching a peak value, corresponding to the critical resolved shear stress (CRSS). The configurations shown in **Figs. 4a3-4c3**, just before CRSS, reveal distinct bowing behaviors. In system A1+, the free dislocation exhibits pronounced bowing; in B1+, the free dislocation remains nearly straight while driving forward bowing of GB dislocations; and in C2+, both the free and LAGB dislocations bow forward. These differences reflect the dependence of bowing behavior on dislocation character—edge dislocations tend to bow themselves, whereas screw or screw-like dislocations favor driving the GB dislocations to bow.

In the transmission stage, the stored elastic stress drives the free dislocation to successively overcome two GB dislocations, leading to a nearly linear stress drop in stress for B1+ and C2+ systems, whereas A1+ shows an initial drop followed by a short plateau and then a linear decrease. Examination of the configurations in **Fig. 4a4** reveals that the plateau begins with depinning from the <001> junction and ends with recombination of collinear annihilation, indicating that collinear annihilation imposes a stronger pinning effect. By contrast, the two GB dislocations are of the same type; thus, once the free dislocation detaches from the first GB dislocation, the second provides little additional resistance (**Figs. 4b4-4c4**), resulting in a continuous linear stress drop. After the sequential detachment from both GB dislocations, the free dislocation fully transmitted across the LAGB and glides toward the free surface (**Figs. 4a5-4c5**). Its eventual absorption at the surface terminates plastic stress relaxation, after which the stress rises linearly again with increasing strain.

we systematically examined the 22 dislocation-LAGB interaction types summarized in **Table 1**, considering both ~2° and ~4° LAGBs for comparison. Each simulation was repeated five times with different random initial velocities. To quantify the resistance of LAGBs to free dislocations, we measured the CRSS for each interaction type and calculated the net transmission stress by subtracting the flow stress of the corresponding free dislocation. The results are summarized in **Fig. 5** (with detailed CRSS and flow stress values provided in **Figs. S2 and S3** of the Supplementary Information).

As shown in **Fig. 5**, the net transmission stress of ~4° LAGBs is approximately twice that of ~2° LAGBs, owing to the nearly doubled GB dislocation density in ~4° LAGBs, which provides more GB dislocations hindering the motion of free dislocations. Consequently, the resistance of LAGBs increases with increasing misorientation. The sign of the Burgers vector has very limited effect on the net transmission stress. For (1$\bar{1}$0) LAGBs, different interaction types exhibit similar net transmission stress, whereas for (111) LAGBs the net transmission stress varies substantially among interaction types. Overall, (010) LAGBs provide the strongest resistance, which is attributed to the strong ternary junction reaction between 1/2<111> and <001> dislocations that significantly hinders dislocation transmission.

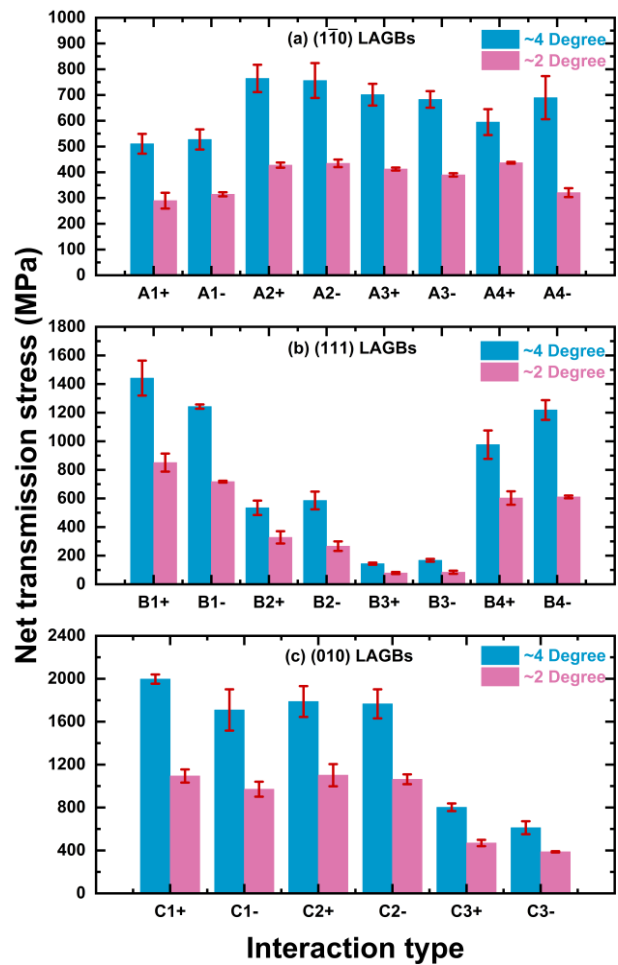

**Fig. 5.** Net transmission stresses of free dislocations across LAGBs for all interaction types defined in **Table 1**. The net value is obtained by subtracting the average flow stress of the corresponding free dislocation from the CRSS, thereby excluding the intrinsic resistance of free dislocations. (a) (1$\bar{1}$0) LAGBs, (b) (111) LAGBs, and (c) (010) LAGBs. Each value represents the average of five independent simulations, and the error bars denote their standard deviation.

# 4. Discussion

The above results reveal a substantial variation in the net transmission stress of free dislocations through LAGBs, with the highest and lowest net transmission stresses differing by nearly an order of magnitude. While previous studies[21, 22] have predominantly attributed this barrier effect to dislocation reaction products (e.g., collinear annihilation, binary <001> junction, and ternary 1/2<111> junction), our results indicate that the elastic interaction between free dislocations and GB dislocations plays a more decisive role.

## 4.1 The dominating factor affecting dislocation transmission through LAGBs

To explore the dominant factors governing dislocation transmission across LAGBs, we first analyzed, following the conventional approach, the influence of dislocation junction length on the resistance to transmission. The post-reaction structures were further quantified in terms of dislocation junction length (**Figs. 6a-6c**). Based on the average junction length, the reactions can be broadly classified into favorable (>~10 Å) and unfavorable (<~10 Å) junction types, corresponding to energetically favorable and less favorable products, respectively. For favorable-junction reactions, the ~2° models consistently produce significantly longer junctions than the ~4° models, with the junction length in the ~2° models being 1.8-2.1 times that in the ~4° models, whereas unfavorable-junction reactions show nearly identical junction lengths in the two models.

Under the assumption that junctions form along the intersection line of the slip planes of the reacting dislocation[37-39], the influence of the Burgers vector sign on junction length depends critically on the slip-plane relation. When the junction lies on the same slip plane as either the GB dislocation or the free dislocation (A1, A2, B1, B3, C2, and C3), the junction length is essentially insensitive to the Burgers vector sign. By contrast, when the junction does not share a slip plane with either reacting dislocation (A3, A4, B2, B4, and C1), the junction length becomes strongly sign dependent, often switching between favorable and unfavorable products. The only exception is type C1, where junctions of opposite sign exhibit nearly identical lengths, and in the C1+ case the junction deviates from the expected intersection line of the slip planes of the reacting dislocations (**Supplementary Fig. S1i**), implying the involvement of additional mechanisms.

Dislocation products formed through reactions, such as collinear annihilation, binary <001> junctions, and ternary 1/2<111> junctions, are generally considered the primary obstacles to dislocation transmission across LAGBs[21, 22, 24-27]. Under this assumption, longer and more stable junctions would be expected to correlate positively with higher net transmission stresses. To test this, we examined the relationship between junction length and net transmission stress of each dislocation-LAGB interaction. As shown in **Figs. 6d-6f**, no significant correlation is observed between net transmission stress and junction length for the (1$\bar{1}$0), (111), and (010) LAGBs. This suggests that junction length alone cannot account for the barrier effect, and that other mechanisms, such as elastic interactions with GB dislocations, may play a more important role.

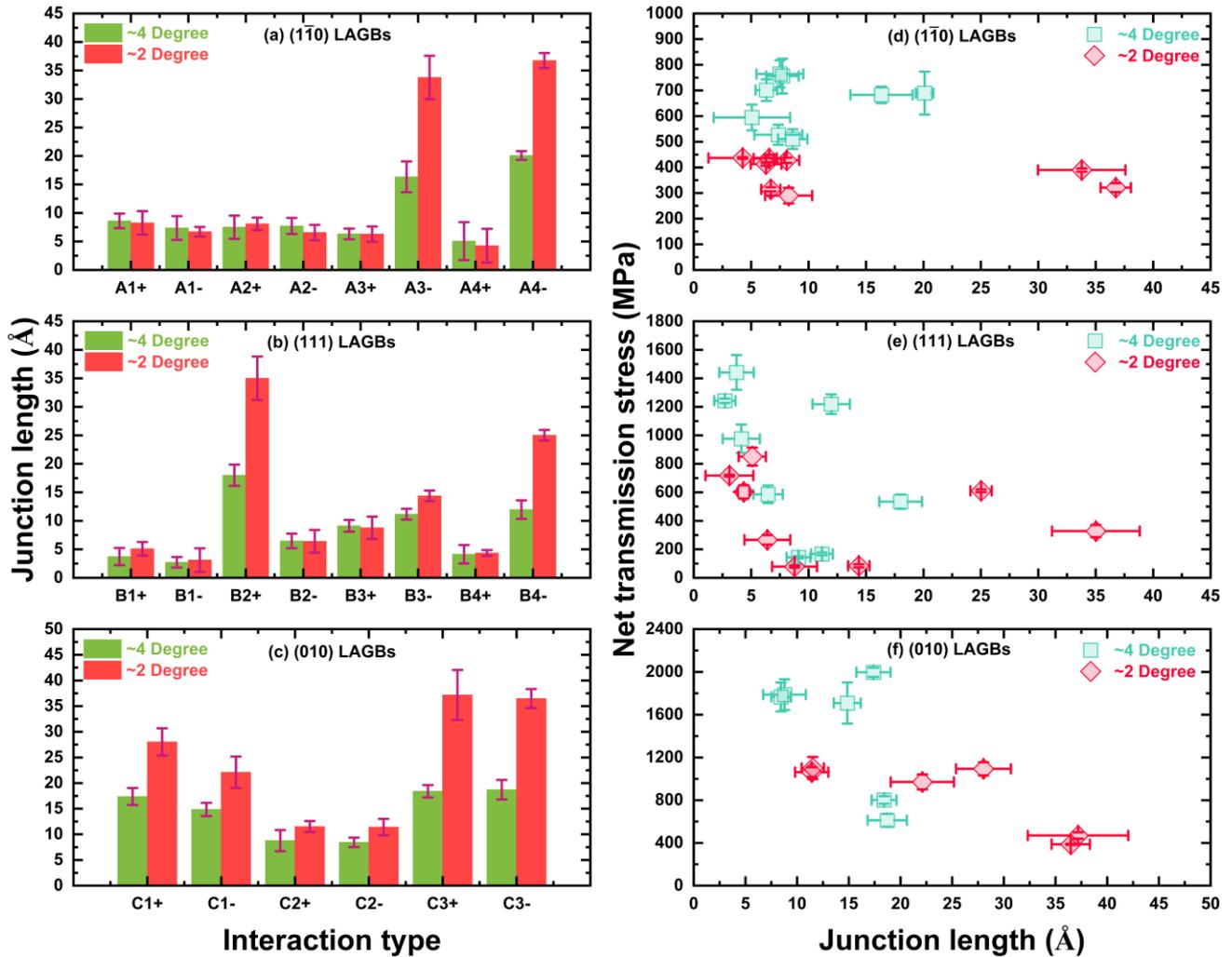

**Fig. 6.** Quantitative analysis of dislocation junctions. (a-c) Average junction lengths; (d-f) correlation between the net transmission stress of free dislocations and junction length. Panels a and d correspond to (1$\bar{1}$0) LAGBs, b and e to (111) LAGBs, and c and f to (010) LAGBs. For junction lengths of (1$\bar{1}$0) LAGBs, only the junction-forming GB dislocation is considered due to collinear annihilation of the other; for those of (111) and (010) LAGBs, values are averaged over two GB dislocations. Data represent the mean of five independent simulations, with error bars denoting their standard deviations.

Apart from the well expected dislocation reactions, a salient feature observed is the pronounced bending of GB dislocations upon interaction with free dislocations. As shown in **Fig. 7**, comparison with their initially straight configurations (blue dashed lines) reveals significant curvature after interaction. This bending elongates the GB dislocation line and thereby raises its elastic energy, a seemingly counterintuitive response that reveals nontrivial atomic-scale reorganization during dislocation-LAGB interactions. A close examination of **Fig. 7** reveal that the degree of bending is closely related to the character angle of the free dislocation, with screw-like dislocations (smaller character angles α) result in more pronounced bending, while edge dislocations (α=90°) induce only minor deviations (additional details on the strong correlation between α and GB bending is given in **Fig. S5** in supplementary information). Interestingly, when comparing the bending behavior in **Fig. 7** with the net transmission stress in **Fig. 5**, we find that severely bent GB dislocations are usually associated with high net transmission stresses (e.g., A2+, B1+, C2+), whereas largely undistorted GBs correspond to relatively weak obstacles (e.g., A1+, B3+, C3+), indicating the pronounced bending of GB dislocations appears to be the key factor controlling the barrier strength of LAGBs.

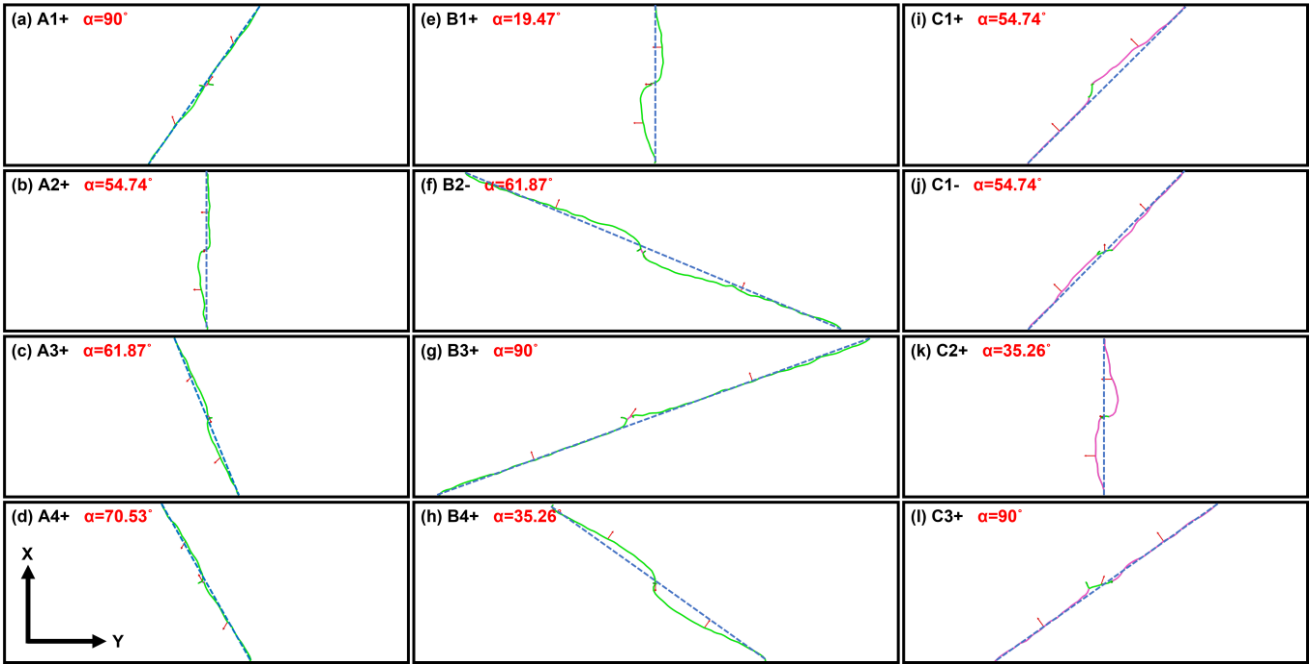

**Fig. 7.** Bending of GB dislocations after interaction with free dislocations. Front views of representative GB dislocations in dislocation-LAGB models with ~2° tilt angle. Blue dashed lines mark the initial GB dislocation positions before interaction. Each panel is labeled by the interaction type (notations follow **Table 1**) and the free dislocation character angle α. (a-d) $(1\bar{1}0)$ LAGBs; (e-h) (111) LAGBs; (i-l) (010) LAGBs. Dislocation coloring is consistent with **Fig. 2**.

To further quantify the bending behavior of GB dislocations, we calculated the ratio of post- to pre-reaction GB dislocation length (**Figs. 8a-8c**), with larger ratios indicating stronger bending. Since junctions are shared segments between free and GB dislocations, two cases were considered: including and excluding junction lengths. Both methods yielded consistent trends, confirming the robustness of the results, and do not alter the overall conclusions. Both approaches reliably capture GB dislocation bending, except for a few types with very long junctions (A3−, A4−, B2+, and B4−). We then examined the relationship between GB dislocation bending and net transmission stress (**Figs. 8d-8f**). A clear positive correlation emerges, demonstrating that greater bending strengthens the barrier effect of

LAGBs on free dislocation transmission. Similar trends are observed when junction contributions are excluded (**Supplementary Fig. S6**), further supporting this conclusion. Although several interactions involving exceptionally long junctions introduce certain scattering (solid points in **Figs. 8d-8e**), the overall positive correlation remains robust. Taken together, these results establish GB dislocation bending as the dominant mechanism governing the barrier effect of LAGBs against free dislocation transmission.

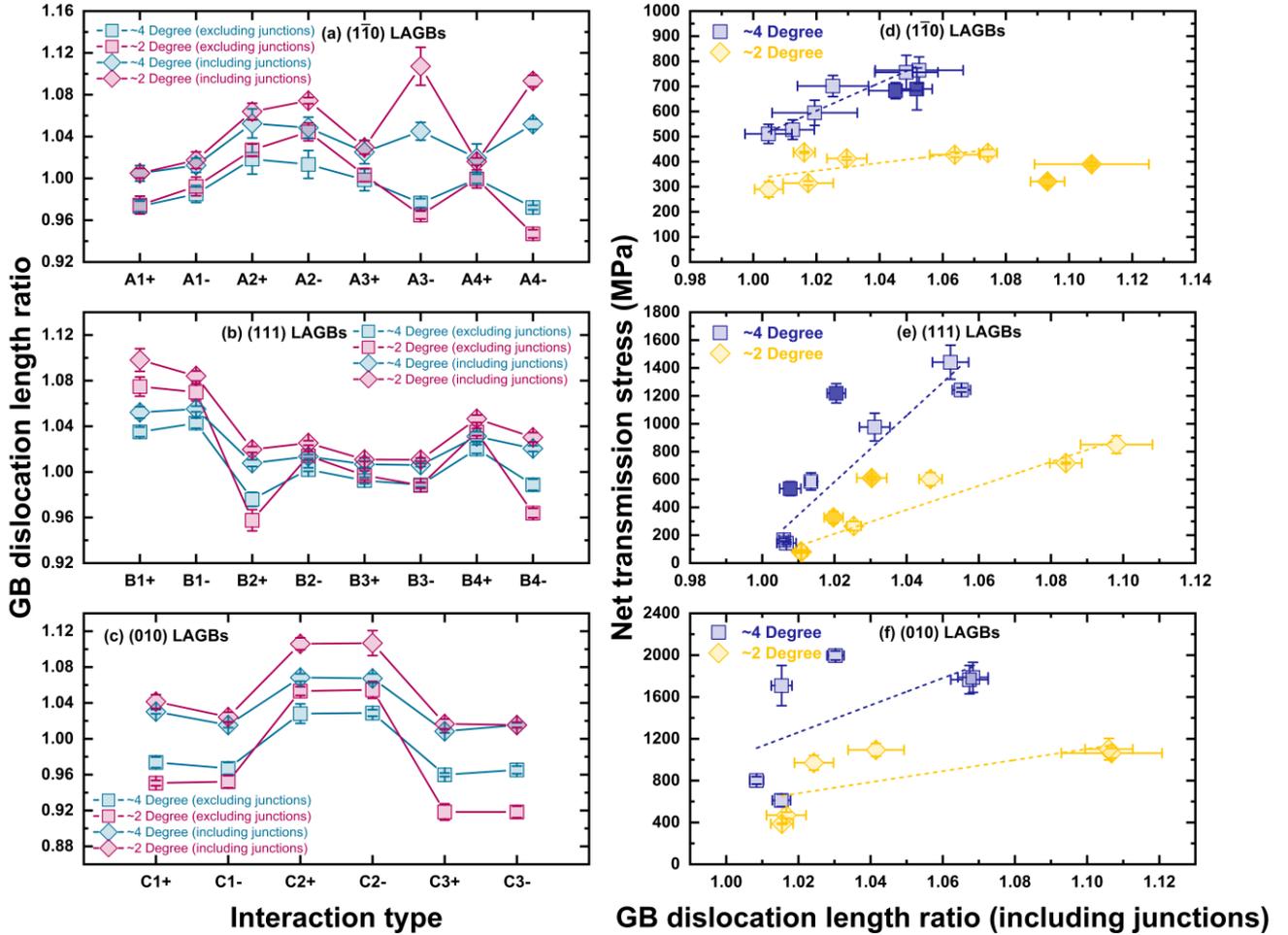

**Fig. 8.** Quantification of GB dislocation bending. (a-c) Bending of GB dislocation for different interaction types (notations follow **Table 1**), defined as the ratio of post- to pre-reaction GB dislocation length, calculated with junctions included and excluded; (d-f) relation between net transmission stress of free dislocations and GB dislocation bending; solid symbols indicate interaction types forming long junctions (A3–, A4–, B2+, and B4–; notations follow **Table 1**). Data represent averages of five independent simulations, with error bars showing standard deviations.

## 4.2 Elastic origin and quantitative prediction of the net transmission stress

The above discussion established the key role of GB dislocation bending in impeding free dislocation transmission, yet the factor controlling the bending magnitude remains unresolved. As shown in **Fig. 7**, GB dislocations tend to maintain straight configurations to minimize line energy, implying that the observed bending must originate from long-range elastic interactions with approaching free dislocations. To elucidate the underlying mechanism, we evaluated the elastic force exerted by a free dislocation on a GB dislocation. Consider an edge-type GB dislocation with its slip plane aligned to the XY plane: it primarily responds to shear stresses $\tau_{zx}$ or $\tau_{zy}$. These shear components are present in the stress field of a screw-type free dislocation lying along the Z axis, but absent for an edge-type free dislocation, which produces only $\tau_{xy}$. Consequently, a strong elastic interaction is expected between GB dislocations and screw-type free dislocations, whereas edge-type dislocations should induce negligible bending. This prediction is fully consistent with the atomistic observations in **Fig. 7**, where screw-like free dislocations impose much stronger GB bending compared to their edge counterparts.

To further quantify the elastic force on GB dislocations, we calculated the stress distribution around the screw free dislocation aligned along the Z direction, the shear stresses are given by[9]:

$$\tau_{zy} = \frac{Gb}{2\pi} \frac{x}{(x^2 + y^2)}, \tag{3}$$

where G is the shear modulus and b is the Burgers vector of the free dislocation. The distribution of $\tau_{zy}$ shown in **Fig. 9a** exhibits a sign reversal across the X= 0 plane is observed, which drives opposite bending of adjacent segments of an edge-type GB dislocation. In more general cases where the GB plane is inclined at an angle β to the YZ plane, the shear stress component $\tau_{zb}$ symmetrically equivalent to $\tau_{zy}$, except rotated by $(\frac{\pi}{2} - \beta)$. Based on this stress distribution, we performed a line scan of the resolved shear stress $\tau_{zb}$ along the GB dislocation line, also considering free dislocations with different character angles α. Mixed dislocations were treated as a superposition of their screw and edge components. As shown in **Fig. 9b**, the magnitude of $\tau_{zb}$ decreases systematically with increasing α, approaching zero at α = 90°, where the elastic driving force for bending vanishes. This analysis rationalizes the observation in **Fig. 7**, where screw-like free dislocations (with small α) impose strong bending on GB dislocations, whereas edge dislocations (α = 90°) produce negligible bending, thereby linking dislocation character directly to the barrier effect of LAGBs.

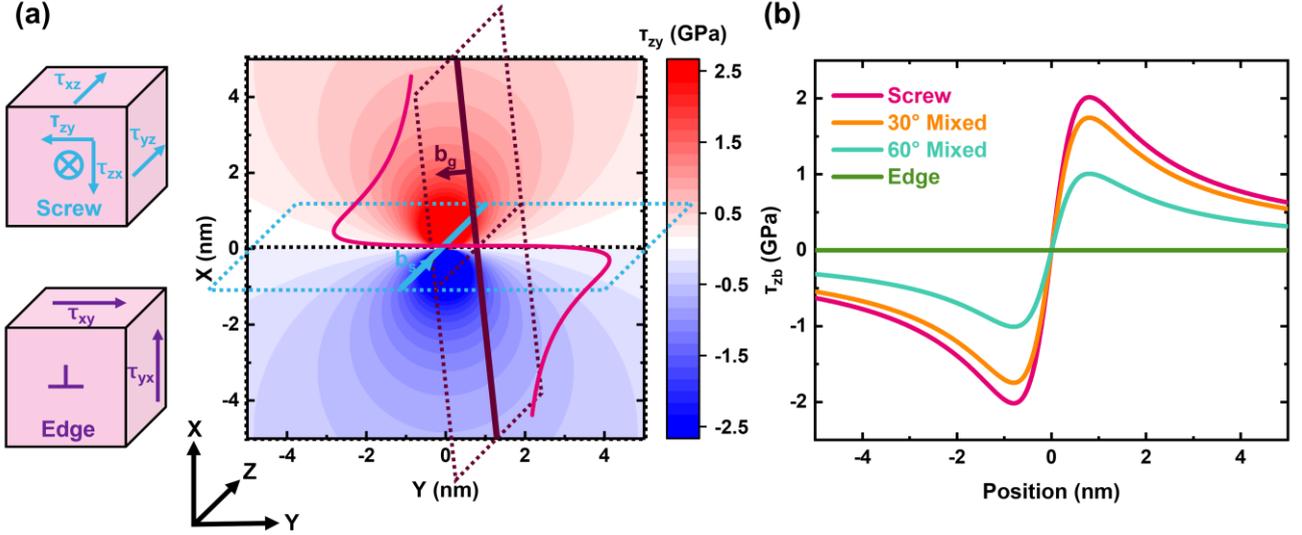

**Fig. 9.** Elastic interaction between free and GB dislocations. (a) Schematic of shear stress fields generated by screw and edge dislocations lying along the Z direction (left). The contour map (right) shows the $\tau_{zy}$ distribution of a screw free dislocation (blue solid line; Burgers vector $b_s$) on the slip plane of GB dislocation (dark red solid line, Burgers vector $b_g$). The magenta solid line indicates the resolved shear stress $\tau_{zb}$ sampled along the GB dislocation line, with b denoting the direction of $b_g$. (b) Variation of $\tau_{zb}$ along the GB dislocation line for free dislocations with different character angles α.

Based on the analysis in **Fig. 9**, we identify three key factors governing the net transmission stress of a dislocation across a LAGB. First, the screw component of the free dislocation, proportional to $\cos\alpha$, dictates the magnitude of shear stresses capable of driving opposite glide of GB dislocation segments. Second, the inclination angle $\beta$ between the LAGB plane and the slip plane modulates the projection of this interaction force, scaling with $\sin\beta$. Third, the misorientation $\theta$ determines the overall density of GB dislocations, thereby setting the baseline resistance through pinning density which scales with $\sin\theta$. Together, these factors yield the following relation:

$$\tau_{nt} = \sin\theta(\tau_e\cos\alpha\sin\beta + \tau_r), \tag{4}$$

where $\tau_e$ and $\tau_r$ are model parameters that respectively correlates with elastic interaction and non-elastic interactions (e.g. dislocation reactions), with fitted values summarized in **Table 3**. Remarkably, **Eq. (4)** reproduces the MD results with excellent agreement (**Fig. 10**), establishing a quantitative framework for predicting dislocation transmission across LAGBs. Among three groups of LAGBs, the (1$\bar{1}$0) LAGBs exhibits the smallest $\tau_e$, indicating a relatively weak dependence of barrier strength on the dislocation character (i.e., elastic interaction). This can be attributed to the fact that half of the GB dislocations undergo colinear annihilation with the free dislocation, leaving only the other half to experience significant elastic interaction. In contrast, the (111) and (010) LAGBs show larger $\tau_e$ values, suggesting a stronger sensitivity of their barrier strength to elastic interaction which depends on the dislocation character.

Among the three LAGBs, the highest $\tau_r$ is observed for (010) LAGBs, reflecting the strong resistance imposed by ternary 1/2<111> junctions. The (1$\bar{1}$0) LAGBs exhibit an intermediate $\tau_r$, consistent with the dominance of collinear annihilation. The lowest $\tau_r$ appears in (111) LAGBs, owing to the formation of binary <001> junctions with comparatively weak resistance. Accordingly, the resistance hierarchy of the three reactions is established as ternary 1/2<111> junction > collinear annihilation > binary <001> junction. More broadly, **Eq. (4)** provides a unified quantitative framework that link between dislocation character and GB geometry, thereby enabling predictive evaluation of LAGB-mediated strengthening in a wide range of crystalline systems. This elastic interaction mechanism, together with the quantitative correlation established in **Eq. (4)**, provides a clear physical explanation for the experimental[20] and simulation[24] observations that LAGBs impose stronger barriers on screw dislocations than on edge dislocations. Moreover, by incorporating the term $\tau_r$, the model also captures the additional contribution of dislocation reactions, reconciling earlier reports[21, 22, 24] that emphasized their role in dislocation transmission. Considering the combined effects of dislocation character, GB geometry, and reaction pathways, the present framework successfully rationalizes prior experimental observations[19] that LAGBs act as "tunable" obstacles to dislocation motion, and provides a predictive foundation for designing materials strengthened by LAGBs.

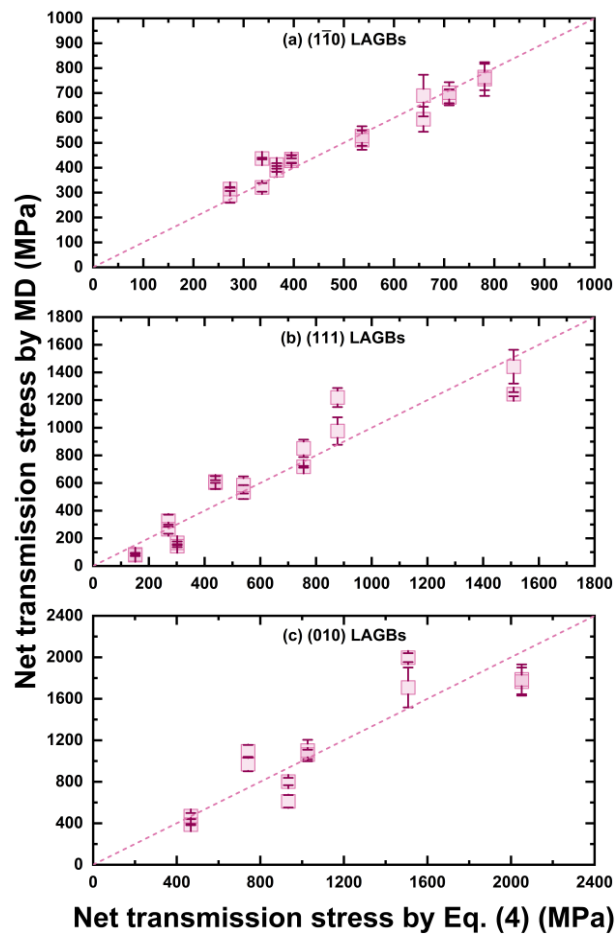

**Fig. 10.** Comparison between the net transmission stress predicted by **Eq. (4)** and MD results for (a) (1$\bar{1}$0), (b) (111), and (c) (010) LAGBs.

**Table 3.** Model parameters ($\tau_e$ and $\tau_r$) for **Eq. (4)**.

| LAGB | $\tau_e$(MPa) | $\tau_r$(MPa) |
|---|---|---|
| ($1\bar{1}0$) LAGBs | 6142.38 | 7784.59 |
| (111) LAGBs | 17959.53 | 4440.52 |
| (010) LAGBs | 19375.76 | 13224.68 |

Despite these advances, the present study focuses on tilt-type LAGBs, which are primarily composed of edge GB dislocations. In contrast, the mechanisms governing dislocation interaction and transmission across twist-type LAGBs, composed of screw GB dislocations, remain less understood, particularly regarding the role of GB dislocation bending induced by elastic stress fields. Furthermore, our results highlight that restructuring of GB dislocations is central to dislocation-LAGB interactions, yet this process can be strongly inhibited when pinned by solute atoms such as hydrogen[40], thereby potentially altering the barrier effect of LAGBs. Addressing these open issues in future studies is crucial for building a comprehensive atomic-scale framework of LAGB-dislocation interactions and for advancing the design of high-performance LAGB-strengthened materials.

## 5. Conclusion

In this work, large-scale MD simulations were employed to elucidate the mechanisms by which tilt LAGBs impede dislocation transmission in BCC Fe. The results reveal that LAGBs exhibit widely variable resistance, acting as either strong or weak barriers depending on geometry and orientation relation between boundary and dislocation. Contrary to the prevailing emphasis on reaction products such as junctions, elastic interactions between free and boundary dislocations emerge as the dominant source of resistance. Screw and screw-like dislocations generate shear stresses that drive opposite glide of GB segments, producing pronounced bending and strong barriers, whereas edge dislocations fail to induce such motion and therefore transmit more readily. The net transmission stress thus increases with decreasing dislocation character angle, with screw dislocations encountering the strongest resistance.

Building on this mechanism, an analytical model was developed that quantitatively predicts the net transmission stress as a function of dislocation character, LAGB inclination and misorientation, in excellent agreement with simulation results. Importantly, this model also provides a quantitative explanation for long-standing experimental and simulation observations that LAGBs block screw dislocations more effectively than edge dislocations. These findings refine the theoretical framework of dislocation-GB interactions and offer new guidance for exploiting LAGBs to strengthen and toughen structural materials.

## Data availability

The data generated and/or analyzed within the current study will be made available upon reasonable request to the authors.

## Acknowledgement

This work was financially supported by National Natural Science Foundation of China (No.: 52401010; 12522517), Hunan Provincial Natural Science Foundation of China (No.: 2025JJ40005). We acknowledge Hefei Advanced Computing Center for providing computing resources.

## Competing interests

The authors declare no competing interests.

## Author contributions

**Shuai Zhang:** Methodology, Formal analysis, Investigation, Data Curation, Writing - Original Draft, Visualization. **Zhishun Chen:** Writing - Review & Editing. **Zhuoming Xie:** Resources, Writing - review & editing. **Jun Song:** Resources, Writing - review & editing. **Huiqiu Deng:** Writing - review & editing. **Wangyu Hu:** Resources, Supervision, Writing - review & editing. **Jie Hou:** Conceptualization, Methodology, Validation, Formal analysis, Resources, Data curation, Writing - Review & Editing, Supervision, Project administration, Funding acquisition.

## Reference

[1] Z.F. Zhang, Z.G. Wang, Y.M. Hu, Fatigue crack initiation and fracture behavior of a copper bicrystal with a perpendicular grain boundary, Materials Science and Engineering: A 269(1-2) (1999) 136-141.

[2] Z.F. Zhang, Z.G. Wang, Comparison of fatigue cracking possibility along large- and low-angle grain boundaries, Mat Sci Eng a-Struct 284(1-2) (2000) 285-291.

[3] Z.F. Zhang, Z.G. Wang, Dependence of intergranular fatigue cracking on the interactions of persistent slip bands with grain boundaries, Acta Materialia 51(2) (2003) 347-364.

[4] B. Cui, J. Kacher, M. McMurtrey, G. Was, I.M. Robertson, Influence of irradiation damage on slip transfer across grain boundaries, Acta Materialia 65 (2014) 150-160.

[5] M.D. McMurtrey, G.S. Was, B. Cui, I. Robertson, L. Smith, D. Farkas, Strain localization at

dislocation channel–grain boundary intersections in irradiated stainless steel, International Journal of Plasticity 56 (2014) 219-231.

[6] L. Wan, W.T. Geng, A. Ishii, J.-P. Du, Q. Mei, N. Ishikawa, H. Kimizuka, S. Ogata, Hydrogen embrittlement controlled by reaction of dislocation with grain boundary in alpha-iron, International Journal of Plasticity 112 (2019) 206-219.

[7] R. Li, K. Wang, W. Zhu, S. Xiao, X. Li, S. Yao, The effects of Escaig stress and solution concentration on the interaction between screw dislocation and coherent twin boundary in random alloys, Computational Materials Science 235 (2024).

[8] Y. Cheng, M. Mrovec, P. Gumbsch, Atomistic simulations of interactions between the $1/2\langle 111\rangle$ edge dislocation and symmetric tilt grain boundaries in tungsten, Philosophical Magazine 88(4) (2008) 547-560.

[9] D. Hull, D.J. Bacon, Introduction to Dislocations, Elsevier2011.

[10] W. Xu, Y.C. Xin, B. Zhang, X.Y. Li, Stress corrosion cracking resistant nanostructured Al-Mg alloy with low angle grain boundaries, Acta Materialia 225 (2022).

[11] Z.C. Tang, W. Xu, D.Y. Zhao, B. Zhang, Improving the strength and SCC resistance of an Al-5Mg-3Zn alloy with low-angle grain boundary structure, Journal of Materials Science & Technology 161 (2023) 63-73.

[12] X.F. Xie, Z.M. Xie, R. Liu, Q.F. Fang, C.S. Liu, W.-Z. Han, X. Wu, Hierarchical microstructures enabled excellent low-temperature strength-ductility synergy in bulk pure tungsten, Acta Materialia 228 (2022).

[13] S. Morito, H. Tanaka, R. Konishi, T. Furuhara, T. Maki, The morphology and crystallography of lath martensite in Fe-C alloys, Acta Materialia 51(6) (2003) 1789-1799.

[14] S. Morito, X. Huang, T. Furuhara, T. Maki, N. Hansen, The morphology and crystallography of lath martensite in alloy steels, Acta Materialia 54(19) (2006) 5323-5331.

[15] A. Nagao, C.D. Smith, M. Dadfarnia, P. Sofronis, I.M. Robertson, The role of hydrogen in hydrogen embrittlement fracture of lath martensitic steel, Acta Materialia 60(13-14) (2012) 5182-5189.

[16] A. Nagao, C.D. Smith, M. Dadfarnia, P. Sofronis, I.M. Robertson, Interpretation of Hydrogen-induced Fracture Surface Morphologies for Lath Martensitic Steel, Procedia Materials Science 3 (2014) 1700-1705.

[17] N. Kheradmand, A.F. Knorr, M. Marx, Y. Deng, Microscopic incompatibility controlling plastic deformation of bicrystals, Acta Materialia 106 (2016) 219-228.

[18] A. Nagao, M. Dadfarnia, B.P. Somerday, P. Sofronis, R.O. Ritchie, Hydrogen-enhanced-plasticity mediated decohesion for hydrogen-induced intergranular and "quasi-cleavage" fracture of lath martensitic steels, Journal of the Mechanics and Physics of Solids 112 (2018) 403-430.

[19] S. Chen, Q. Yu, The role of low angle grain boundary in deformation of titanium and its size effect, Scripta Materialia 163 (2019) 148-151.

[20] S. Ueki, S. Morito, Anisotropic slip behaviour of lath martensite block in ultra-low carbon steel, Scripta Materialia 255 (2025).

[21] B. Liu, D. Raabe, P. Eisenlohr, F. Roters, A. Arsenlis, G. Hommes, Dislocation interactions and low-angle grain boundary strengthening, Acta Materialia 59(19) (2011) 7125-7134.

[22] B. Liu, P. Eisenlohr, F. Roters, D. Raabe, Simulation of dislocation penetration through a general low-angle grain boundary, Acta Materialia 60(13-14) (2012) 5380-5390.

[23] S. Kondo, T. Mitsuma, N. Shibata, Y. Ikuhara, Direct observation of individual dislocation interaction processes with grain boundaries, Sci Adv 2(11) (2016) e1501926.

[24] M. Wakeda, T. Ohmura, Atomistic evaluation of the dislocation transmission across tilt and twist low-angle grain boundaries in body-centered cubic iron, Computational Materials Science 228 (2023).

[25] N. Verdhan, R. Kapoor, Interaction of dislocations with low angle tilt boundaries in fcc crystals, Computational Materials Science 98 (2015) 149-157.

[26] R. Kapoor, N. Verdhan, Interaction of dislocation pile-up with a low-angle tilt boundary: a discrete dislocation dynamics study, Philosophical Magazine 97(7) (2016) 465-488.

[27] Y. Gao, Z. Jin, Interactions between lattice dislocation and Lomer-type low-angle grain boundary in nickel, Computational Materials Science 138 (2017) 225-235.

[28] Y.T. Chou, Dislocation reactions and networks in anisotropic b.c.c. crystals, Materials Science and Engineering 10 (1972) 81-86.

[29] S. Plimpton, Fast Parallel Algorithms for Short-Range Molecular Dynamics, Journal of Computational Physics 117(1) (1995) 1-19.

[30] A. Stukowski, Visualization and analysis of atomistic simulation data with OVITO–the Open Visualization Tool, Modelling and Simulation in Materials Science and Engineering 18(1) (2010).

[31] A. Stukowski, V.V. Bulatov, A. Arsenlis, Automated identification and indexing of dislocations in crystal interfaces, Modelling and Simulation in Materials Science and Engineering 20(8) (2012).

[32] G.J. Ackland, M.I. Mendelev, D.J. Srolovitz, S. Han, A.V. Barashev, Development of an interatomic potential for phosphorus impurities in α-iron, Journal of Physics: Condensed Matter 16(27) (2004) S2629-S2642.

[33] M.I. Mendelev, S. Han, D.J. Srolovitz, G.J. Ackland, D.Y. Sun, M. Asta, Development of new interatomic potentials appropriate for crystalline and liquid iron, Philosophical Magazine 83(35) (2003) 3977-3994.

[34] P. Hirel, Atomsk: A tool for manipulating and converting atomic data files, Computer Physics Communications 197 (2015) 212-219.

[35] J.-Y. Zhang, W.-Z. Zhang, A general method to construct dislocations in atomistic simulations, Modelling and Simulation in Materials Science and Engineering 27(3) (2019).

[36] K.T. Kashyap, A. Bhat, P.G. Koppad, K.B. Puneeth, On Peierls Nabarro stress in Iron, Computational Materials Science 56 (2012) 172-173.

[37] L.K. Wickham, K.W. Schwarz, J.S. Stölken, Rules for Forest Interactions between Dislocations, Physical Review Letters 83(22) (1999) 4574-4577.

[38] R. Madec, B. Devincre, L.P. Kubin, On the nature of attractive dislocation crossed states, Computational Materials Science 23(1-4) (2002) 219-224.

[39] S.M. Hafez Haghighat, R. Schäublin, D. Raabe, Atomistic simulation of the a0 <100> binary junction formation and its unzipping in body-centered cubic iron, Acta Materialia 64 (2014) 24-32.

[40] J. Hou, D. Peng, X.-S. Kong, H. Deng, W. Hu, C. Chen, J. Song, Hydrogen Modulated Dislocation Reaction and Defect Accumulation in bcc Metals, Acta Materialia (2025) 121524.

# Supplementary information for Dislocation Transmission Across Tilt Low-Angle Grain Boundaries in BCC Fe: The Role of Elastic Interactions

## S1. Three-view analysis of post-reaction dislocation structures

Supplementary **Fig. S1** provides three-view images of representative post-reaction configurations. The front views correspond to the GB dislocations shown in **Fig. 7**, whereas the other two views provide additional details of the free dislocations and reaction products. The main reaction products include binary <001> junctions, collinear annihilation, and ternary 1/2<111> junctions. Specifically, 1/2<111> free dislocations interacting with (1$\bar{1}$0) LAGBs form a <001> junction with one GB dislocation and undergo collinear annihilation with another; reactions with (111) LAGBs produce two binary <001> junctions; and reactions with (010) LAGBs result in two ternary 1/2<111> junctions involving <001> GB dislocations.

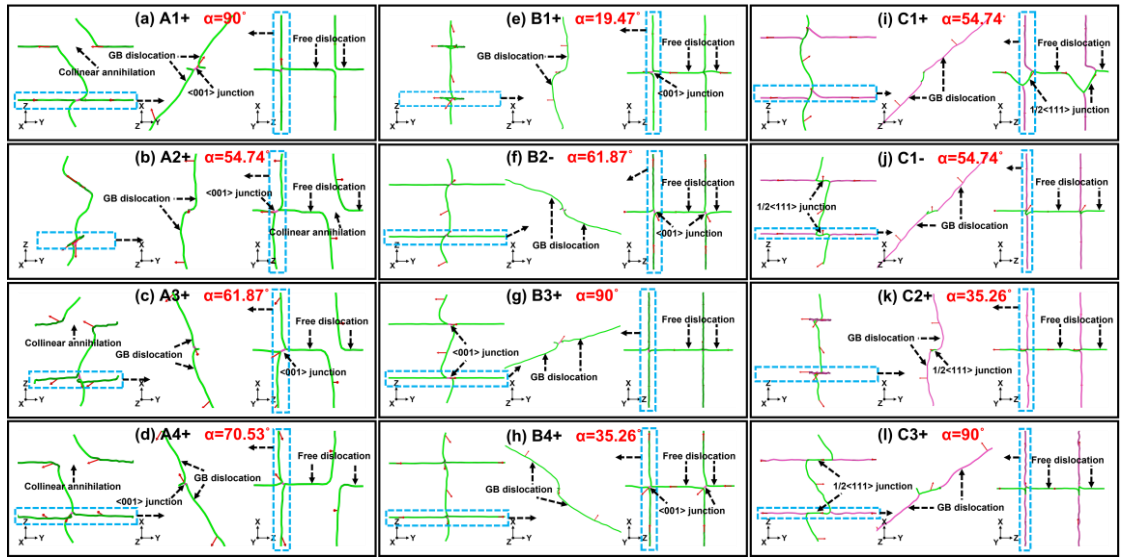

**Figure S1.** Three-view images (top, front, and right) of representative post-reaction dislocation structures for dislocation-LAGB interaction models with ~2° tilt angle. The front view corresponds to the GB dislocation indicated by the blue dashed rectangle in the other two views, identical to the GB dislocation shown in **Fig. 7**. Each panel is labeled by the interaction type (notations follow **Table 1**) and the free dislocation character angle α. (a-d) (1$\bar{1}$0) LAGBs; (e-h) (111) LAGBs; (i-l) (010) LAGBs. Dislocation coloring follows **Fig. 2**.

## S2. CRSS for all interaction types

**Fig. S2** summarizes the CRSS for all interaction types, representing the minimum stress required for a free dislocation to transmit across an LAGB. This stress includes both the resistance from the LAGB and the intrinsic resistance of the free dislocation. Because the intrinsic resistance is much smaller than that of the LAGB and the average flow stress varies by less than 70 MPa among different free dislocations, the CRSS in **Fig. S2** shows trends similar to those of the net transmission stress in **Fig. 5**, leading to consistent relative magnitudes of critical stresses.

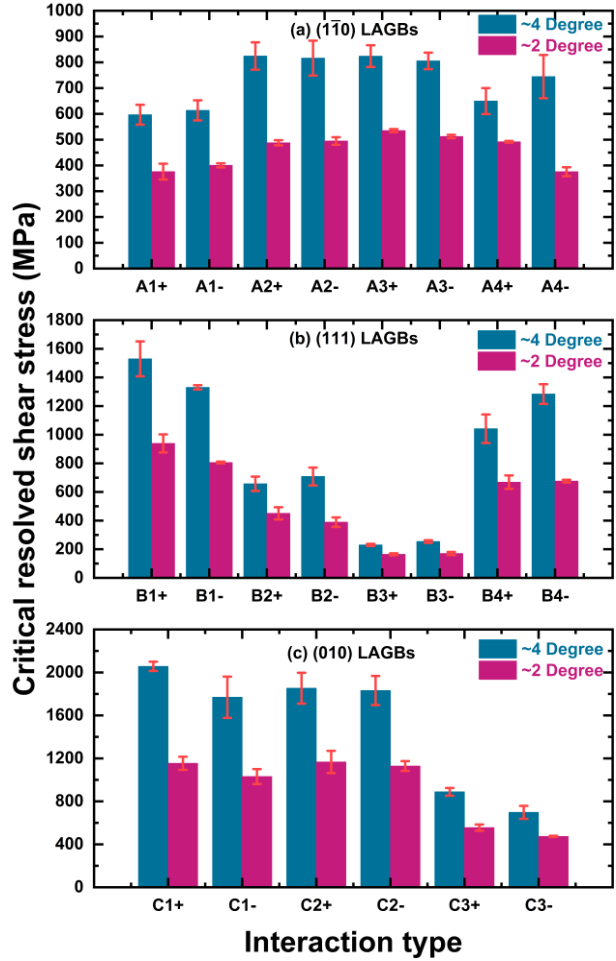

**Figure S2.** CRSS for the transmission of free dislocations across LAGBs for all interaction types defined in **Table 1**. (a) (1$\bar{1}$0) LAGBs, (b) (111) LAGBs, and (c) (010) LAGBs. Each value represents the average of five independent simulations, and the error bars denote their standard deviation.

## S3. Flow stress of free dislocations

To evaluate the intrinsic resistance of free dislocations, the steady-state flow stress of all dislocation types involved in the dislocation-LAGB interaction models was measured under identical loading conditions (300 K, shear strain rate 0.05/ns). In contrast to the interaction models, periodic boundary conditions were also applied along the dislocation motion direction (Y) to allow for continuous dislocation glide and ensure steady-state motion.

The evolution of the resolved shear stress $\tau_{xb}$ during loading is shown in **Fig. S3a**. All dislocation types exhibited similar behavior: $\tau_{xb}$ increased approximately linearly with MD steps, reached a peak, and then fluctuated around a constant value, indicating steady-state glide. The average $\tau_{xb}$ over $2\times10^5$-$1\times10^6$ MD steps was taken as the flow stress, which ranged from 50 to 130 MPa for different dislocation types (**Fig. S3b**).

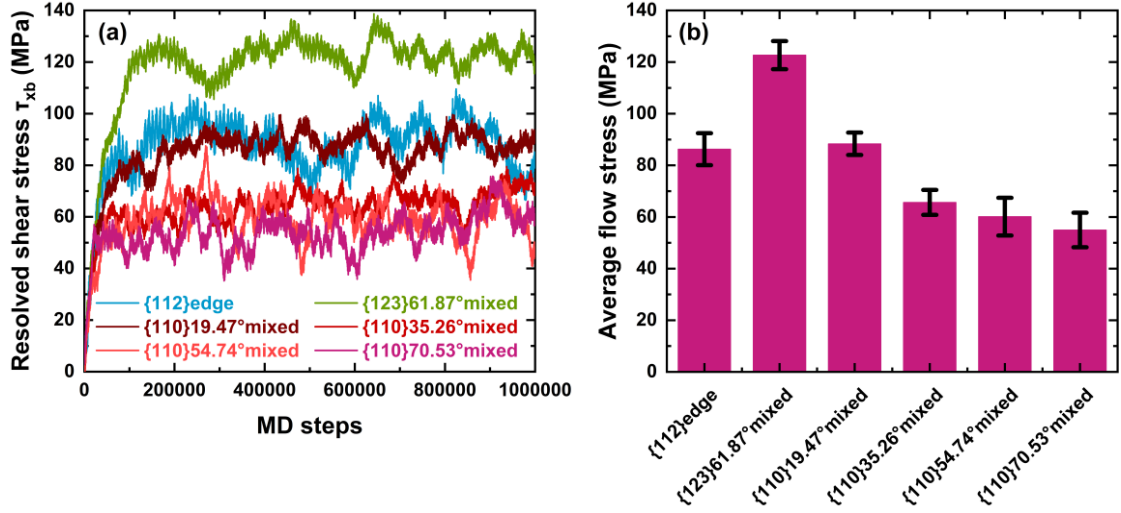

**Figure S3.** Flow stress of free dislocations. (a) Evolution of resolved shear stress $\tau_{xb}$ with MD steps, where x and b denote the slip-plane normal and the Burgers vector, respectively. (b) Average flow stress over $2\times10^5$-$1\times10^6$ MD steps; error bars indicate standard deviations.

## S4. Effects of line length and loading direction on flow stress

To confirm the reliability of the obtained flow stresses, the effects of dislocation line length and loading direction were examined. The line-length effect was tested using {112} edge and {110} 19.47° mixed dislocations (the latter being closest to a screw character), with line lengths corresponding to ~2° and ~4° dislocation-LAGB models. As shown in Supplementary **Fig. S4a**, the influence of line length on flow stress was negligible. In addition, considering that both positive and negative Burgers vectors were involved in the interaction models, corresponding to opposite loading directions, three types of dislocations ({112} edge, {110} 19.47° mixed, and {123} 61.87° mixed) were further tested. The loading direction showed no significant influence on the flow stress (**Fig. S4b**).

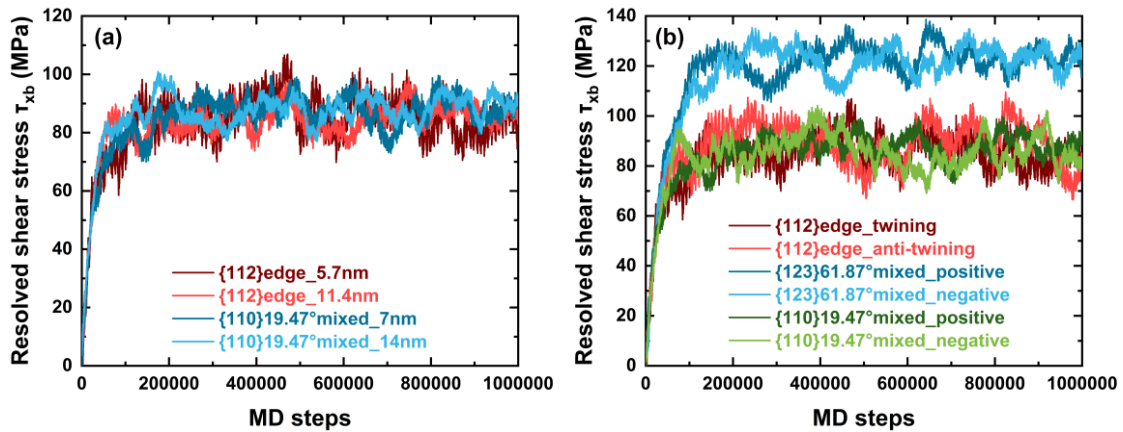

**Figure S4.** Effects of dislocation line length and loading direction on flow stress. (a) Evolution of resolved shear stress $\tau_{xb}$ for line lengths corresponding to ~4° and ~2° dislocation-LAGB models. (b) $\tau_{xb}$ evolution under different loading directions (twinning/anti-twinning, positive/negative).

## S5. GB dislocation bending as a function of cos α

In the main text (**Fig. 8**), GB dislocation bending was quantified as the ratio of the post- to pre-reaction GB dislocation length. Larger cos α (i.e., closer to screw character) leads to more pronounced GB dislocation bending. It should be noted that the solid symbols in **Fig. S5** (A3–, A4–, B2+, and B4–, denoted as in **Table 1**) involve certain deviations due to the formation of long dislocation junctions.

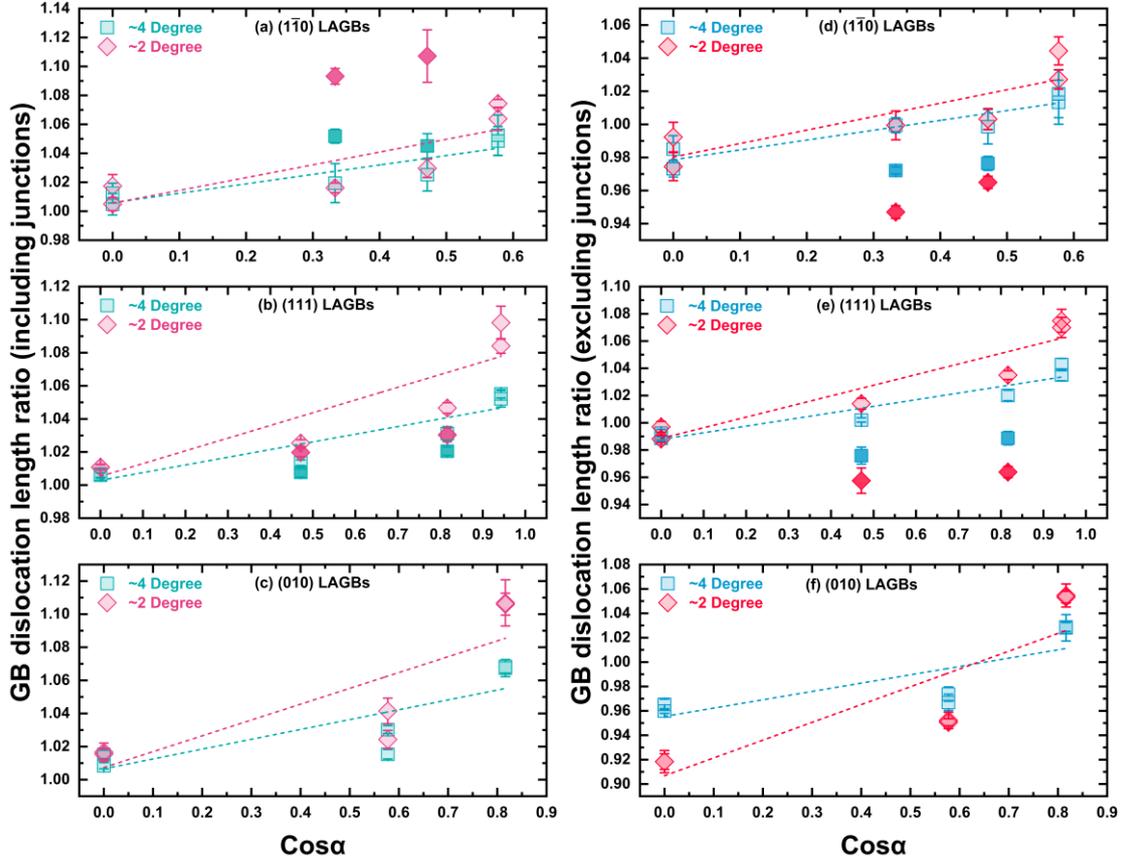

**Figure S5.** Dependence of GB dislocation bending on free dislocation character angle α. (a-c) Bending with junctions included; (d-f) bending with junctions excluded. Panels a and d correspond to (1$\bar{1}$0) LAGBs, b and e to (111) LAGBs, and c and f to (010) LAGBs. Bending is defined as the ratio of post- to pre-reaction GB dislocation length. Solid symbols highlight interaction types forming long junctions (A3–, A4–, B2+, B4–; notations follow **Table 1**). Data represent the mean of five independent simulations, with error bars denoting their standard deviations.

## S6. Correlation between net transmission stress and GB dislocation bending, with junction length excluded

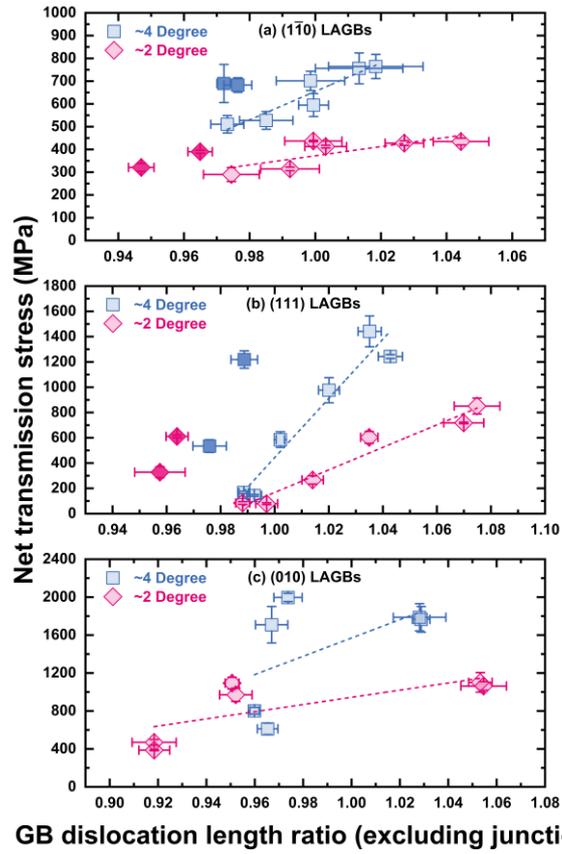

**Figure S6.** Net transmission stress vs. GB dislocation bending. (a) (1$\bar{1}$0), (b) (111), and (c) (010) LAGBs. Bending is defined as the ratio of post- to pre-reaction GB dislocation length, excluding junctions. Solid symbols highlight cases forming long junctions (A3–, A4–, B2+, and B4–; notations follow **Table 1**). Data represent the mean of five independent simulations, with error bars denoting their standard deviations.